\documentclass[12pt]{article}
\usepackage{graphicx}
\usepackage[cp1251]{inputenc}
\begin{document}
\pagestyle{empty}

\pagestyle{headings} {\title{Solar flares and their associated
processes}}
\author{O.M.Boyarkin$^a$\thanks{E-mail:boyarkin@front.ru},
\ I.O.Boyarkina\thanks{E-mail:estel20@mail.ru}\ $^b$\\
$^a$\small{\it{Belarusian State University,}}\\
\small{\it{Dolgobrodskaya Street 23, Minsk, 220070, Belarus}}\\
$^b$\small{\it{University of Rome Tor Vergata}}\\
\small{\it{Via Orazio Raimondo 18, Roma 00173 Lazio, Rome,
Italy}}}
\date{}
\maketitle
\begin{abstract}

Consideration is being given to the neutrino system governed by
the wave function $\Psi^T=(\nu_{eL},\nu_{XL}, \overline{\nu}_{eL},
\overline{\nu}_{XL})$ traveling in the region of the solar flare
(SF). Our treatment of the problem holds for any standard model
(SM) extensions in which massive neutrinos possess nonzero dipole
magnetic and anapole moments. The possible resonance conversions
of the electron neutrinos are examined. Since the $\nu_{eL}\to
\nu_{XL}$ and $\nu_{eL}\to \overline{\nu}_{XL}$-resonances take
place in the convective zone, their existence can in no way be
connected with the SF. However, when the solar neutrino flux moves
through the SF region in the preflare period, then it undergoes
the additional resonance conversions resulting in appearance of
the $\overline{\nu}_{eL}$ and $\overline{\nu}_{XL}$ neutrinos. On
the other hand, according to the hypothesis of the $\nu_e$-induced
$\beta$-decays weakening the electron neutrinos flux leads to the
decrease of the $\beta$-decay rate for some elements. So, under
passage of the electron neutrino flux through the region of the
SF, the following four phenomena could be detected: (i) decreasing
the number of the electron neutrinos; (ii) appearance of the
$\overline{\nu}_{eL}$ neutrinos; (iii) increasing the amount of
the $\overline{\nu}_{XL}$-neutrinos; (iv) reduction of the
$\beta$-decay rate for some elements of the periodic table.

The possible influence of the electron antineutrino flux produced
in the superflares on the regime of the hypothetical georeactor is
considered. \\[1mm]

PACS number(s): 12.60.Cn, 14.60.Pg, 96.60.Kx, 95.85.Qx, 96.60.Rd.
\end{abstract}

Keys words: Solar flares, neutrino, dipole magnetic moment,
anapole moment, neutrino telescopes, resonance transitions,
$\nu_e$-induced $\beta$-decays, georeactor.

\section{Introduction}

At certain conditions the evolution of active regions (ARs) on the
Sun may lead to the appearance of solar flares (SFs), which
represent the most powerful events of the solar activity. The
energy generated in the process of the SF is about of
$10^{28}-10^{32}$ erg. However, as it was shown in Ref.
\cite{ML17}, the superflares whose energy could be as large as
$10^{36}$ are also possible. It is believed that just the magnetic
field provides a main energy source of the SF's. The commonly
accepted model of the SF production is the magnetic reconnection
model which is based on breaking and reconnection of magnetic
field strength lines of neighboring spots. According to it the
process of the SF evolution is as follows \cite{KST11}. The SF
formation starts from integrating sunspots of fairly opposite
polarity. Then changing the magnetic field configuration could
result in the appearance of a limiting strength line (LSL) being
common for whole group. Throughout the LSL which rises from
photosphere to the corona the redistribution of magnetic fluxes
happens. From the moment of the LSL appearance an electric field
induced by magnetic field variations causes current along this
line. By virtue of the interaction with a magnetic field this
current takes a form of a current layer (CL). Because the CL
prevents from the magnetic fluxes redistribution, the process of
magnetic energy storage of the CL begins. Duration of the
formation period of the CL (the initial SF phase) varies from
several to dozens of hours. The second stage (an explosion phase
of the SF) has a time interval of 1-3 minutes. It starts from the
appearance of a high-resistance region in some part of the CL that
results in a current dissipation. Then, due to penetration of the
magnetic field through the CL a strong magnetic field appears
perpendicular to it. An arising magnetic force breaks the CL and
throws out a plasma at a great speed. The magnetic energy of
sunspots is transformed into kinetic energy of matter emission (at
a speed of the order of $10^6\ \mbox{m/s}$), energy of hard
electromagnetic radiation, and fluxes of solar cosmic rays which
consist of protons, nuclei with charges $2\leq {\mbox{z}}\leq28$,
and electrons. Produced photons reach the Earth by approximately
8.5 minutes after explosion phase of the SF. Further during some
tens of minutes powerful flux of charged particles attains
terrestrial surface. As far as the plasma clouds are concerned,
they reach our planet within two-three days only. The flux falling
on the Earth's surface for the most powerful the SF may reach
$\sim 4500\%$ in comparison to the background flux of cosmic
particles. The concluding stage (the hot phase of the SF) is
characterized by the existence of a high temperature coronal
region and can continue for several hours. One of the
characteristic futures of flares is their isomorphism, that is,
the repetition in one and the same place with the same field
configuration. A small flare may repeat up to 10 times per day
while a large one may take place the next day and even several
times during the active region lifetime.

It is clear that the high-power SF's can be especially destructive
when they proves to be aimed at the Earth. They are dangerous for
satellites, astronauts, pilots, power grids and so on. Moreover,
intensive fluxes of the electron antineutrinos produced during the
SF could influence on the work of a georeactor. It is obvious that
forecasting the SF at the initial phase is a very important task.

In this work we are going to investigate the phenomena which are
connected with the SF's. In the next section we consider the
evolution of the neutrino flux in the solar matter and magnetic
field. Our treatment of the problem carries rather general
character, namely, it holds for any standard model extensions in
which neutrinos have masses and possess both the magnetic dipole
and anapole moments with the values being close to the current
experimental bounds. We find all possible the resonance
conversions of the electron neutrino flux which travels the region
of the SF in preflare phase. Section 3 is devoted to our
conclusions.

\section{Solar neutrino flux}

Our goal is to obtain an evolution of the neutrino system
traveling through the region of the SF. So, we have to discuss
both the neutrino electromagnetic properties and characteristics
of the solar magnetic field.

Let us start with the neutrino multipole moments (MM's). Neutrinos
are neutral particles and their total Lagrangian does not contains
any MM's. These moments are caused by the radiative corrections.
The most general form of the matrix element for the conserved
neutrino electromagnetic current $J_{\mu}^{em}$ could be obtained
from the demands of the relativistic and gauge invariance. It is
easy to show that for a Dirac neutrino $(J_{\mu})_{if}$ must have
the form
$$<\nu^D_i(p^{\prime})|J_{\mu}^{em}|\nu^D_j(p)>=<\nu^D_i(p^{\prime})|i
\sigma_{\mu\lambda} q^{\lambda} [F_M(q^2)+F_E(q^2)\gamma_5]+$$
$$+(q^2\gamma_{\mu}-q_{\mu}\hat{q})[F_V(q^2)+F_A(q^2)\gamma_5]
 |\nu^D_j(p)>,\eqno(1)$$ where $q=p^{\prime}-p$,
$F_M(q^2)$, $F_E(q^2)$, $F_A(q^2)$ and $F_V(q^2)$ are magnetic,
electric, anapole and reduced Dirac formfactors. In static limit
($q^2=0$) $F_M(0)$ and $F_E(0)$ define dipole magnetic moment
(DMM) $\mu_{ij}$ and dipole electric moment $d_{ij}$,
respectively. At $i=j$, $F_A(0)$ represents the anapole moment
(AM).

For a Majorana neutrino from the $CPT$ invariance it is evident
that all the formfactors, except the axial one $F_A$, are
identically equal to zero. As regards non-diagonal elements, the
situation depends on the fact whether $CP$-parity is conserved or
not. For the $CP$ non-variant case all the four formfactors are
nonzero. When $CP$ invariance takes place and the $|\nu^M_i>$ and
$|\nu^M_j>$ states have identical (opposite) $CP$-parities, then
$(F_E)_{ij}$ and $(F_A)_{ij}$ ($(F_M)_{ij}$ and $(F_V)_{ij}$) are
different from zero.

Now we address the recent experimental bounds on the neutrino DMM
and AM. The world best limit on electron neutrino DMM was derived
from the GEMMA experiment at the Kalinin nuclear power plant
\cite{AGB12}
$$\mu_{\nu_e}\leq2.9\times10^{-11}\mu_B\qquad\mbox{at}\ 90\%\ \mbox{C.L.}
\eqno(2)$$
A search for the solar neutrino effective magnetic moment has been
performed using data from 1291.5 days exposure during the second
phase of the Borexino experiment. The obtained bound is as follows
\cite{MAG17}
$$\mu_{\nu}^{eff}\leq2.8\times10^{-11} \mu_B\qquad \mbox{at}
\ 90\% \ \mbox{C.L.}\eqno(3)$$ Note that the minimally extended SM
(the SM with the massive neutrinos
--- MESM) predicts that the values of the neutrino DMMs are
negligibly small and are of no physical interest \cite{BWLe77}.
Further we remember that at neutrino mass neglecting the AM is
associated with a neutrino charge radius by the relation
$$a_{\nu_l}={1\over6}<r_{\nu_l}^2>.\eqno(4)$$
Measuring the neutral-current coherent elastic neutrino-nucleus
scattering at the TEXONO experiment leads to the following bounds
\cite{TSK15}
$$\mu_{\nu_e}\leq0.83\times10^{-10}\mu_B,\qquad
-0.17\times10^{-32}\ \mbox{cm}^2\leq<r_{\nu_e}^2>\leq
0.17\times10^{-32}\ \mbox{cm}^2.\eqno(5)$$ On the other hand,
using the lowest-energy solar neutrino data of $pp, ^7Be$ and
$pep$ spectra from phase-I and phase-II runs of Borexino
experiment results in \cite{ANK17}
$$\mu_{\nu}\leq8.75\times10^{-12}\mu_B,\qquad
-0.82\times10^{-32}\mbox{cm}^2\leq<r_{\nu_e}^2>\leq
1.27\times10^{-32}\ \mbox{cm}^2.\eqno(6)$$

Proceeding to the solar magnetic field we note that the field
strength $B$ of a big sunspots ($d\sim2\times10^5$ km) could reach
$10^4$ Gs while their geometrical depth $h$ is approximately 300
km. In this case a magnetic field above sunspots is characterized
by geometrical phase $\Phi(z)$
$$B_x\pm iB_y = B_{\bot}e^{\pm i\Phi(z)}\eqno(7)$$
and has non-potential character
$$(\mbox{rot}\ {\bf{B}})_z=4\pi j_z.\eqno(8)$$
The data concerning centimeter radiation above a spot is
indicative of a gas heating up to the temperatures of a coronal
order. For example, at the height $\sim 2\cdot 10^2$ km the
temperature could be as large as $10^6$ K, that leads to a great
value of solar plasma conductivity ($\sigma\sim T^{3/2}$). That
permits to assume, that the density of longitudinal electric
current might be large enough in a region above a spot. For
example, in Ref. \cite{KDA15} it was shown that AR's current could
reach the values of $(0.7-4)\times10^{12}$ A.

In our consideration we shall be limited by two generations.
Therefore, the subject of our consideration will be a neutrino
system consisting of $\nu_{eL}, \nu_{XL}$ and their anti-particles
$(\nu_{eL})^c, (\nu_{XL})^c$ ($c$ means an operation of charge
conjugation, and $X=\mu,\tau$). It should be stressed that at a
switching on of weak interaction the Majorana neutrino does not
represent an eigenstate of a charge conjugation operator.  Then,
by virtue of the fact that $(\nu_{eL})^c$ and $(\nu_{eX})^c$ are
right-handed neutrinos, we shall employs for them both in Majorana
and Dirac cases following designations ${\overline\nu}_{eL}$ and
${\overline\nu}_{XL}$ respectively.

For the system of the Dirac neutrinos the evolution equation will
look like
$$i\frac{d}{dz}\left(\matrix{\nu_{eL}\cr
\nu_{XL}\cr\overline{\nu}_{eL}\cr\overline{\nu}_{XL}\cr}\right)
={\cal H}\left(\matrix{\nu_{eL}\cr
\nu_{XL}\cr\overline{\nu}_{eL}\cr\overline{\nu}_{XL}\cr}\right)
,\eqno(9)$$ where
$${\cal H}=$$
$$ \left(\matrix{
           -\delta^{12}_c+V_{eL}+4\pi a_{\nu_e\nu_e}j_z &
           \delta^{12}_s+4\pi a_{\nu_e\nu_X}j_z &
\mu_{\nu_e\overline{\nu}_e} B_{\bot}e^{i\Phi} & \mu_{\nu_e
\overline{\nu}_X}B_{\bot}e^{i\Phi}\cr
          \delta^{12}_s+4\pi a_{\nu_X\nu_e}j_z
           &\delta^{12}_c+V_{XL}+
           4\pi a_{\nu_X\nu_X}j_z &
\mu_{\nu_e\overline{\nu}_X}B_{\bot}e^{i\Phi} & \mu_{\nu_X
\overline{\nu}_X}B_{\bot}e^{i\Phi}\cr \mu_{\nu_e\overline{\nu}_e}
B_{\bot}e^{-i\Phi} & \mu_{\nu_e
\overline{\nu}_X}B_{\bot}e^{-i\Phi}&
           -\delta^{12}_c-4\pi a_{\overline{\nu}_e\overline{\nu}_e}j_z&
           \delta^{12}_s-4\pi a_{\overline{\nu}_e\overline{\nu}_X}j_z\cr
\mu_{\nu_e\overline{\nu}_X} B_{\bot}e^{-i\Phi} & \mu_{\nu_X
\overline{\nu}_X}B_{\bot}e^{-i\Phi}&
           \delta^{12}_s-4\pi a_{\overline{\nu}_X\overline{\nu}_e}j_z &
           \delta^{12}_c-4\pi a_{\overline{\nu}_X\overline{\nu}_X}j_z\cr}
           \right),$$
$$\delta^{12}_{c(s)}=\frac{m^2_1-m^2_2}{4E}\cos{2\theta_{\nu}}
(\sin{2\theta_{\nu}}),\qquad V_{eL}=\sqrt{2}G_F(n_e-n_n/2),\qquad
V_{XL}=-\sqrt{2}G_Fn_n/2,$$ $n_e$ and $n_n$ are electron and
neutron densities, respectively, $\theta_{\nu}$ is a mixing angle
in vacuum between mass eigenstates $\nu_1$ and $\nu_2$, $V_{eL}$
($V_{XL}$) is a matter potential describing interaction of the
$\nu_{eL}$ ($\nu_{XL}$) neutrino with a solar matter, and
$a_{\nu_l\nu_{l^{\prime}}}$ is an anapole moment between $\nu_l$
and $\nu_{l^{\prime}}$ states.

When the neutrino has a Majorana nature, we must set
$$\mu_{\nu_e\overline{\nu}_e}=\mu_{\nu_X\overline{\nu}_X}=0,\eqno(10)$$
and replace \ $-4\pi a_{\overline{\nu}_l \overline{\nu}_l}j_z$ \
($l=e,X$)\ on \ $4\pi
a_{\overline{\nu}_l\overline{\nu}_l}j_z-V_l$.

Having carried out a phase rotation
$$\left(\matrix{\nu_{eL}^{\prime}\cr
\nu_{XL}^{\prime}\cr\overline{\nu}_{eL}^{\prime}\cr\overline{\nu}
_{XL}^{\prime}\cr}\right)=S\left(\matrix{\nu_{eL}\cr
\nu_{XL}\cr\overline{\nu}_{eL}\cr\overline{\nu}_{XL}\cr}\right),\eqno(11)$$
where $$\left(\matrix{e^{i\Phi/2}&
           0 & 0 & 0\cr
0 & e^{i\Phi/2}& 0 & 0\cr 0&0& e^{-i\Phi/2} & 0\cr 0&0&0&
e^{-i\Phi/2}}\right),$$ and $\Psi^{\prime T}=(\nu_{eL}^{\prime},
\nu_{XL}^{\prime},\overline{\nu}_{eL}^{\prime},\overline{\nu}
_{XL}^{\prime})$ is the vector state of neutrino system in the
reference frame rotating with the same angular velocity as the
transverse magnetic field.  The expression for the transformed
Hamiltonian ${\cal{H}}^{\prime}=S{\cal{H}}S^{-1}$ follows from the
old one by the substitutions
$$\left.\begin{array}{ll}e^{\pm i\Phi}\to1,\qquad {\cal{H}}_{11}^{\prime}\to
{\cal{H}}_{11}- \dot{\Phi}/2, \qquad
{\cal{H}}_{22}^{\prime}\to{\cal{H}}_{22}-\dot{\Phi}/2,\\[2mm]
\hspace{14mm}{\cal{H}}_{33}^{\prime}\to{\cal{H}}_{33}+\dot{\Phi}/2,
\qquad
{\cal{H}}_{44}^{\prime}\to{\cal{H}}_{44}+\dot{\Phi}/2\end{array}\right\}
\eqno(12)$$ By virtue of the fact that
$|\Psi^{\prime}|^2=|\Psi|^2$ we shall drop the prime sign in what
follows.

Further we shall concentrate on the resonance conversions of the
electron neutrinos. When the neutrinos are the Dirac particles
there are three resonance conversions. The $\nu_{eL}\to\nu_{XL}$
(Micheev-Smirnov-Wolfenstein --- MSW) resonance  is the first. It
is realized when
$$\Sigma_{\nu_{eL}\to\nu_{XL}}=
-2\delta^{12}_c+V_{eL}-V_{XL}+4\pi (a_{\nu_e\nu_e}-
a_{\nu_X\nu_X})j_z=0.\eqno(13)$$ while the transition width is as
follows
$$\Gamma(\nu_{eL}\to\nu_{XL})\simeq
{\sqrt{2}(\delta_s^{12}+4\pi a_{\nu_e\nu_X}j_z)\over
G_F}.\eqno(14)$$ Attention is drawn to the fact that both
$\Sigma_{\nu_{eL}\to\nu_{XL}}$ and $\Gamma(\nu_{eL}\to\nu_{XL})$
depend on the neutrino energy. As a result in this resonance only
electron neutrinos with the energy of order of few MeV take part.
It is also believed to be well established that the MSW resonance
may occur before the convective zone, that is, it takes place
whether the SF happens or not.

The second one is the $\nu_{eL}\to{\overline\nu}_{XL}$ resonance
happening with flavor and spin flipping. It occurs at the
condition
$$\Sigma_{\nu_{eL}\to\overline{\nu}_{XL}}=-2\delta^{12}_c+V_{eL}+4\pi
(a_{\nu_e\nu_e}+a_{\overline{\nu}_X\overline{\nu}_X})j_z-\dot{\Phi}=0.
\eqno(15)$$ Corresponding expressions for the transition width
will look like
$$\Gamma(\nu_{eL}\to\overline{\nu}_{XL})\simeq
{\sqrt{2}\mu_{\nu_e\overline{\nu}_X}B_{\perp}\over
G_F}.\eqno(16)$$ Comparing Eqs.(13) and (15) we see that the
$\nu_{eL}\to\overline{\nu}_{XL}$ resonance occurs at a higher
density than the MSW one. It should be stressed that since the
transition width of the MSW resonance does not depend on the DMM
then it proves to be allowed within the SM. As far as the
$\nu_{eL}\to\overline{\nu}_{XL}$ resonance is concerned, then its
realization is possible only in the model with nonzero DMM. The
$\nu_{eL}\to\overline{\nu}_{XL}$ resonance may occur not only in
the convective zone, but also in the chromosphere where the
neutrino MM's may play an important role. However, in the latter
case this resonance could take place only when
$\dot{\Phi}\sim10^{-7}\ \mbox{cm}^{-1}$ what is improbable. This
seemingly circumstance leads to the conclusion that in the solar
neutrino flux which travels the SF region at the preflare period
the amount of $\overline{\nu}_{XL}$ neutrinos appearing under
$\nu_{eL}\to\overline{\nu}_{XL}$ resonance transition will not be
changed. However it is not the case. The $\nu_{XL}$ neutrinos
produced under the MSW resonance could undergo the resonance
transition $\nu_{XL}\to\overline{\nu}_{XL}$ that takes place when
the condition
$$\Sigma_{\nu_{XL}\to\overline{\nu}_{XL}}=
V_{XL}+4\pi(a_{\nu_X\nu_X}+a_{\overline{\nu}_X\overline{\nu}_X})j_z-
\dot{\Phi}=0\eqno(17)$$ is realized. The corresponding transition
width is defined as
$$\Gamma(\nu_{XL}\to\overline{\nu}_{XL})\simeq
{\sqrt{2}\mu_{\nu_X\overline{\nu}_X}B_{\perp}\over
G_F}.\eqno(18)$$ We note, that notwithstanding the fact, that both
$\Sigma_{\nu_{XL}\to\overline{\nu}_{XL}}$ and
$\Gamma(\nu_{XL}\to\overline{\nu}_{XL})$ do not depend on the
energy, only neutrinos having energy of order of few MeV exhibit
this resonance conversion.

The third resonance conversion is
$\nu_{eL}\longrightarrow{\overline\nu}_{eL}$ It will proceed when
the condition
$$\Sigma_{\nu_{eL}\to\overline{\nu}_{eL}}=V_{eL}+4\pi(a_{\nu_e\nu_e}
+a_{\overline{\nu}_e \overline{\nu}_e})j_z-\dot{\Phi}=0\eqno(19)$$
is carried out. In this case the transition width has the view
$$\Gamma(\nu_{eL}\to\overline{\nu}_{eL})\simeq
{\sqrt{2}\mu_{\nu_e\overline{\nu}_e}B_{\perp}\over
G_F}.\eqno(20)$$ As we see both
$\Sigma_{\nu_{eL}\to\overline{\nu}_{eL}}$ and
$\Gamma(\nu_{eL}\to\overline{\nu}_{eL})$ do not display the
dependence on the neutrino energy. This, in its turn, means that
all electron neutrinos produced in the center of the Sun ($pp$-,
$^{13}N$-,...and $hep$-neutrinos) may undergo
$\nu_{eL}\to\overline{\nu}_{eL}$ resonance transitions. We note if
$B_{\perp}\mu_{\nu_e\overline{\nu}_e}\simeq\delta^{12}_s$, then
the transition width of this resonance will have the same order of
magnitude as the MSW one. Since the transition time in the
level-crossing region is proportional to the transition width,
then we can expect that approximately half electron neutrinos are
subjected to the $\nu_{eL}\to\overline{\nu}_{eL}$ conversion. For
example, in that case the flux of the antineutrinos with the
energy up to 0.5 MeV could be as large as $10^{11}\
\mbox{cm}^{-2}\mbox{s}^{-1}.$

Further we assume that $\nu_{eL}\to\overline{\nu}_{eL}$-resonance
take place in chromosphere. Then from Eq. (19) it follows this
resonance will be allowed in two simplest cases: (i)
$$\dot{\Phi}\ll V_{eL}\qquad {\mbox{but}}\qquad
-4\pi(a_{\nu_e\nu_e}+a_{\overline{\nu}_e
\overline{\nu}_e})j_z=V_{eL},\eqno(21)$$ or (ii)
$$4\pi(a_{\nu_e\nu_e}+a_{\overline{\nu}_e
\overline{\nu}_e})j_z\ll V_{eL}\qquad {\mbox{but}}\qquad
\dot{\Phi}=V_{eL}.\eqno(22)$$ To realize requirements (21) the
AM's sum $(a_{\nu_e\nu_e}+a_{\overline{\nu}_e \overline{\nu}_e})$
must have negative sign and the value of $j_z$ must have the order
of $10^{-6}\ \mbox{A}/\mbox{cm}^2$ (where we have used the lower
bound on the AM and the value of the matter potential in
chromosphere $V_{cr}\sim10^{-26}$ eV). On the other hand, for
fulfillment of (22) we need to demand $\dot{\Phi}\sim10^{-21}\
\mbox{cm}^{-1}.$

In the case of Majorana neutrinos only the resonances
$\nu_{eL}\to\nu_{XL}$ and $\nu_{eL}\to\overline{\nu}_{XL}$ are
allowed. The resonance transitions
$\nu_{eL}\to\overline{\nu}_{eL}$ and
$\nu_{XL}\to\overline{\nu}_{XL}$ are absent by reason of zeroth
DMM.

In recent years a number of articles, presenting evidence that
some $\beta$-decay rates are variable, have been published (see,
for up-to-date review, Ref. \cite{TM16}). First result was
presented in Ref.\cite{JHJ2009}. It was observed that the
$\beta$-decay rate of\ $^{54}\mbox{Mn}$ decreased slightly
beginning 39 hours before the large SF of 2006 Dec.13. Later the
hypothesis was offered \cite{TM16} that this changeability may be
connected with decreasing the solar neutrino flux (hypothesis of
the $\nu_e$-induced $\beta$-decays --- H$\nu_e$ID) when it passes
through the region of the SF. Another way of putting it is that
some elements we belief that they are natural radioactive, in
reality, are artificial radioactive because of the solar neutrino
flux bombardment. It should be recalled that for the first time
the idea about correlation of a neutrino flux with the SF's has
been suggested in the works \cite{OBD95}.

It is not inconceivable that the solar electron antineutrinos flux
produced in a super powerful flare could influence on the
operating conditions of a georeactor. For the first time the
georeactor concept was proposed by J.M. Herndon \cite{JMH}. There
are several reasons for its existence. We enumerate the basic
ones. The first reason is connected with the Earth's magnetic
field. It is known that this field varies in intensity and
irregularly reverses polarity with an average interval between
reversals of about $2\times10^5$ years. To ensure that some
variable or intermittent energy source is required. This source is
understood as georeactor, i.e., as naturally varying
self-sustaining nuclear chain reaction burning at the center of
the Earth. The second reason is associated with the attempt of
explanation of increasing the relation $^3\mbox{He}/^4\mbox{He}$
as the distance is varied from the surface to the bottom mantle of
the Earth \cite{DA20}.

In the georeactor $^{239}\mbox{Pu}$ is formed by neutron capture
in $^{238}\mbox{U}$ followed by two short-lived beta decays:
$^{238}\mbox{U}(n,\gamma)\to^{239}\mbox{U}(\beta^-)\to
^{239}\mbox{Np}(\beta^-)\to ^{239}\mbox{Pu}$. The neutron flux in
the reactor is extremely low and, in contrast with man-made high
flux power reactors, $^{239}\mbox{Pu}$ does not contribute to the
fission power and decays in $^{235}\mbox{U}$:
$^{239}\mbox{Pu}(\alpha,T_{1/2}=2.4\times10^4\mbox{y})\to^{235}\mbox{U}$.
Thus the georeactor operates in a breeder regime and reproduces
$^{235}\mbox{U}$ through $^{238}\mbox{U}\to^{239}\mbox{Pu}\to^{
235}\mbox{U}$ cycle. Variations of georeactor power originate from
self-poisoning due to accumulation of fission products and
subsequent removal of these products by diffusion or some other
mechanism. The reactors of such a type are referred to as the
traveling wave reactor. Georeactor numerical simulations indicate
that the georeactor would function as a fast neutron breeder
reactor capable of operating for at least as long as Earth has
existed. Nuclear fission chain reaction can occur in nature. In
1972 such a natural nuclear fission reactor were actually found in
the mine at Oklo in the Republic of Gabon in Africa \cite{MN72}.

At present the searches for the antineutrino flux produced by the
georeactor are carried out in Borexino and KamLAND. Like
investigations will be fulfilled at the following neutrino
telescopes (NT's): SNO+ (Sudbury Neutrino Observatory+)
\cite{SA16}, BNO (Baksan Neutrino Observatory) \cite{IRB17},
HanoHano \cite{DY06} and LENA (Low Energy Neutrino
Astronomy)\cite{MW12} which will be placed into operation in the
nearest future.

It should be stressed that one of the main problems of the
georeactor is connected with the mechanism of the nuclear burning
start-up, that is, with the appearance of the neutron-dividing
wave (see, for example, \cite{VlA00}). One of the possible
explanations is as follows. Let us assume that in the far past the
super solar flare (SSF) has been happened. The ultra high energy
protons injected downwards from the coronal acceleration region
will interact with dense plasma in the lower solar atmosphere
producing mesons which subsequently decay, resulting in, so called
postflare neutrinos with O(MeV-GeV) energies
$$p+p\ \mbox{or} \ p+\alpha\to\left\{\begin{array}{ll}\pi^++X,\qquad
\pi^+\to\mu^++\nu_{\mu},\qquad \mu^+\to
e^++\nu_e+\overline{\nu}_{\mu}\\
 \pi^-+X,\qquad \pi^-\to\mu^-+\overline{\nu}_{\mu},\qquad \mu^-\to
e^-+\overline{\nu}_e+\nu_{\mu}\end{array}\right.\eqno(23)$$
Further one could speculate that the emerged electron antineutrino
flux passing the Earth's thickness faces the region which is
enriched by protons. In this case due to the reaction of the
inverse $\beta^-$ decay the neutrons flux will be born. Then,
these neutrons with the energy $E_n\geq1$ MeV bombarding
$^{238}\mbox{U}$ could cause the nuclear burning start-up of the
georeactor.

\section{Conclusions}
In two flavor approximation the evolution of the solar neutrino
flux traveling the region of the solar flare (SF) has been
investigated. One was assumed that the neutrinos possess both the
dipole magnetic moment (DMM) and the anapole moment. The cases of
the Dirac and Majorana nature of neutrinos have been considered.
The possible resonance conversions of the electron neutrinos have
been examined.

The MSW- and $\nu_{eL}\to \overline{\nu}_{XL}$-resonances take
place in the convective zone, therefore, their existence can in no
way be connected with the SF. However, when the electron neutrinos
pass the SF region in preflare period they are subjected to
additional resonance conversions. In the case of the Majorana
neutrino we may have only the conversion
$\nu_{eL}\to\overline{\nu}_{XL}$ while for Dirac neutrino apart
from that we deal with one more conversion
$\nu_{eL}\to\overline{\nu}_{eL}$. The conditions of the resonances
existence and the transition widths (TW's) have been found. The
TW's of the resonances $\nu_{eL}\to \overline{\nu}_{XL}$,
$\nu_{eL}\to\overline{\nu}_{eL}$, and
$\nu_{XL}\to\overline{\nu}_{XL}$ proves to be proportional to the
DMM. However, the MESM predicts the DMM value close to zero.
Therefore, in the MESM these resonances have the TW's being equal
to zero and, as a result, they must not be observed from the point
of view of this model.

After leaving from the Sun the neutrino flux flies $1,5\times10^8$
km in a vacuum before it will reach the Earth. As this takes
place, reduction of the electron neutrino flux is caused by the
vacuum oscillations which lead solely to $\nu_{eL}\to\nu_{XL}$
transitions. Therefore, when the SF is absent, the NT records the
electron neutrino flux weakened at the expense of vacuum
oscillations, of the MSW resonance, and
$\nu_{eL}\to\overline{\nu}_{XL}$-resonance. However, when the
electron neutrino flux passes the SF region in preflare period
then it is further weakened because of additional resonance
conversions, apart from the above-listed. As the analysis showed
the most probable scenario is the existence of the
$\nu_{eL}\to\overline{\nu}_{eL}$-and
$\nu_{XL}\to\overline{\nu}_{XL}$ -resonances which are allowed for
the Dirac neutrino only. It is worth noting that since both
$\Sigma_{\nu_{eL}\to\overline{\nu}_{eL}}$ and
$\Gamma(\nu_{eL}\to\overline{\nu}_{eL})$ do not depend on the
neutrino energy then all electron neutrino born in the Sun center
may go through the $\nu_{eL}\to\overline{\nu}_{eL}$-resonance.
Having regard to appearing the $\overline{\nu}_{XL}$ neutrinos in
$\nu_{eL}\to\overline{\nu}_{XL}$ transition as well, one may
argue, when the solar neutrino flux moves through the SF region in
the preflare period, the $\overline{\nu}_{eL}$ neutrinos will be
appeared while the amounts of $\overline{\nu}_{XL}$ neutrinos will
be increased. Then, terrestrial detectors could record these
phenomena with the help of the reactions
$$\overline{\nu}_{eL}+p\to n+e^+\eqno(24)$$
$$\overline{\nu}_{\mu L}+p\to n+\mu^+.\eqno(25)$$
Note, the reaction (24) is at the heart of the antineutrino
detectors that are used for nuclear reactor monitoring in the
on-line regime. Since the energy threshold of the reaction (24) is
$\sim1.8$ MeV then the antineutrinos produced from $^{13}N$,
$^{15}O$ and more energetic electron neutrinos could initiate this
reaction. As a result, the value of the electron antineutrinos
flux falling on the proton target of the detector could be as
large as $10^8\ \mbox{cm}^{-2}\mbox{s}^{-1}$.

The hypothesis of the $\nu_e$-induced $\beta$-decays has been
considered as well. According to it the reduction of the
$\beta$-decay rate for some elements in the preflare period is
caused by decreasing the electron neutrinos flux.

So, under passage of the electron neutrino flux through the region
of the SF, the following four phenomena could be detected: (i)
decreasing the number of the electron neutrinos; (ii) appearance
of the $\overline{\nu}_{eL}$ neutrinos; (iii) increasing the
amount of the $\overline{\nu}_{XL}$-neutrinos; (iv) reduction of
the $\beta$-decay rate for some elements of the periodic table. It
should be stressed that all these phenomena will be not only
forerunners of the SF, but they also will be the convincing
arguments in favour both of the Physics beyond the SM and of the
Dirac neutrino nature.

We have also pointed to the possible influence of the electron
antineutrino flux produced in the superflares on the regime of the
hypothetical georeactor.

\section*{Acknowledgments}

This work is partially supported by the grant of Belorussian
Ministry of Education No 20162921.


\begin{thebibliography}{xxxx}
\bibitem{ML17}M. Lingam, A. Loeb, Astrophys. J., {\bf{848}} (2017) 41.
\bibitem{KST11}K. Shibata and T. Magara, Living Rev. Solar Phys.
{\bf{8}} (2011) 6.
\bibitem{AGB12}A. G. Beda {\it{et al.}}, GEMMA Collaboration, Advances in
High Energy Physics {\bf{2012}} (2012) 350150.
\bibitem{MAG17}[Borexino collaboration] M. Agostini {\it{et al.}},
arXiv: 1707.09355, [hep-ph] (2017).
\bibitem{BWLe77}B. W. Lee, R. E. Shrock, Phys. Rev. D {\bf{16}} (1977) 1444.
\bibitem{TSK15}T.S. Kosmas, Phys. Lett. B {\bf{750}} (2015) 459.
\bibitem{ANK17}Amir N. Khan, arXiv: 1709.02930 [hep-ph] (2017).
\bibitem{KDA15}K. Dalmasse {\it{et al.}}, Astrophys. J., {\bf{810}}
(2015) 17.
\bibitem{TM16}T. Mohsinally {\it{et al.}} Astropart. Phys. {\bf{75}} (2016) 29.
\bibitem{JHJ2009}J. H. Jenkins, E. Fischbach, Astropart. Phys.
{\bf{31}} (2009) 407.
\bibitem{OBD95}O. Boyarkin, D. Rein, Zeitschr. Phys. C {\bf{67}} (1995) 607;
O. Boyarkin, Phys. Rev. D {\bf{53}} (1996) 5298; O. Boyarkin,
Russ. Phys. J. {\bf{39}} (1996) 597.
\bibitem{JMH}J. M. Herndon, Proc. Natl. Acad. Sci. USA
{\bf{100}} (2003) 3047; D. F. Hollenbach, J. M. Herndon, ibid.
{\bf{98}} (2001) 11085; J. M. Herndon, J. Geomagn. Geoelectr.
{\bf{45}} (1993) 423.
\bibitem{DA20}D.L. Anderson, Int. Geology Rev. {\bf{42}} (2000)
289.
\bibitem{MN72} M. Neuilly {\it{et al.}}, C. R. Acad. Sci. Paris
D {\bf{275}} (1972) 1847.
\bibitem{SA16}S. Andringa {\it{et al.}}, Adv. High Energy Phys.
2016 {\bf{2016}} (2016) 6194250.
\bibitem{IRB17}I. R. Barabanov {\it{et al.}}, Physics of Atomic
Nuclei {\bf{80}} (2017) 446.
\bibitem{DY06}S. Dye, A Deep Ocean Anti-Neutrino
Detector near Hawaii --- Hanohano, tech. rep., Makai ocean
engineering, inc, University of Hawaii. 09/2006. No. 1228/2009.
\bibitem{MW12}M. Wurm {\it{et al.}}, Astropart. Phys. {\bf{35}} (2012) 685.
\bibitem{VlA00}V. F. Anisichkin {\it{et al.}} Atomnaya Energiya,
{\bf{98}} (2005) 370.
\end{thebibliography}
\end{document}